\documentclass[11pt]{article}
\usepackage{amsfonts}
\usepackage{amssymb}
\usepackage{graphicx}
\usepackage{amsmath}
\usepackage{amsthm}
\usepackage{color}
\usepackage{multirow}
\usepackage{url}
\usepackage{float}
\usepackage{booktabs} 
\usepackage{makecell}
\usepackage{listings}
\usepackage{subcaption}
\usepackage{algorithm}
\usepackage{enumerate}
\usepackage{tikz}

\usepackage{algpseudocode}
\lstnewenvironment{Rcode}{\lstset{language=R}}{}
\newtheorem{theorem}{Theorem}[section]

\newtheorem{corollary}[theorem]{Corollary}

\definecolor{battleshipgrey}{rgb}{0.52, 0.52, 0.51}
\definecolor{navyblue}{rgb}{0.0, 0.0, 0.5}
\definecolor{arsenic}{rgb}{0.23, 0.27, 0.29}
\definecolor{oldmauve}{rgb}{0.4, 0.19, 0.28}
\usepackage[colorlinks=TRUE]{hyperref}
\hypersetup{
	colorlinks=true,
	linkcolor=navyblue,
	filecolor=blue,      
	urlcolor=oldmauve,
	citecolor=navyblue,
	pdftitle={Overleaf Example},
	pdfpagemode=FullScreen,
}
\usepackage{natbib}
\setcitestyle{authoryear}
\begin{document}
	
	\title{ \bf A $k$NN procedure in semiparametric functional data analysis}
	\author{Silvia Novo{$^a$}\footnote{Corresponding author email address: \href{s.novo@udc.es}{s.novo@udc.es}} \hspace{2pt} Germ\'{a}n Aneiros{$^b$} \hspace{2pt} Philippe Vieu{$^c$} \\		
		{\normalsize $^a$ Department of Mathematics, MODES, CITIC, Universidade da Coruña, A Coruña, Spain}\\
		{\normalsize $^b$ Department of Mathematics, MODES, CITIC, ITMATI, Universidade da Coruña, A Coruña, Spain}\\
		{\normalsize $^c$ Institut de Math\'{e}matiques, Universit\'e  Paul Sabatier, Toulouse, France}
	}
	
	\date{}
	\maketitle
	\begin{abstract} \normalsize
	A fast and flexible $k$NN procedure is developed for dealing with a semiparametric functional regression model involving both partial-linear and single-index components. Rates of uniform consistency  are presented. Simulated experiments highlight the advantages of the $k$NN procedure. A real data analysis is also shown.
	\end{abstract}
	\vspace{1cm}
	\noindent \textit{Keywords: } Functional single-index; Partial Linear; $k$NN regression;  Semiparametric FDA.
\section{Introduction}
Functional Data Analysis (FDA) became one among the main topics in the research statistical literature for at least two reasons: on the one hand there are more and more applied scientific fields having to face with functional datasets (see eg \citealt{anecv19} for a recent overview of applied issues in FDA), while on the other hand there are many methodological challenges to front for analyzing such data (see eg  \citealt{anecfv19} for an overview on methodological issues on FDA). The first difficulty when building statistical models is to balance the trade-off between flexibility and dimensionality. Said with other words, a statistical model should be able to reduce dimensionality effects (see eg \citealt{vie19} for discussion), but being still able to capture as wide as possible information on the data. In regression setting, pure nonparametric models (see \citealt{linv18} for a  survey) are highly affected by dimensionality effects while semiparametric ideas are more appealing candidates.

Once a model has been specified, the second important question  is to develop accurate statistical procedures. Such estimation techniques should be as much as possible data-driven because the complexity of functional dataset may make rather difficult any empirical choice of intricated parameters. Moreover, functional problems are involving very big datasets and the statistical procedures should be of fast implementation. In regression settings, $k$-Nearest-Neighbours ($k$NN) ideas have two main advantages: firstly they  depend only on a simple integer parameter (the number of neighbours) making the method fast and easy, and secondly they provide location-adaptive estimates being able to capture local features of the data.

The aim of this paper is to propose a general semiparametric functional regression model allowing for sets of predictors being mixture of functional and multivariate ones. The model (see Section \ref{model}) is combining single-index ideas (for dealing with functional predictor) together with partial-linear ideas (for dealing with multivariate one). Then, we  develop a $k$NN procedure for estimating the smooth components of the model (see Section \ref{estimates}). In Section \ref{theory}, rates of uniform consistency  are obtained in a general  way allowing for fully automatic estimates. As a by-product, we  state similar results for usual Nadaraya-Watson functional kernel regression. A short simulation study is reported along Appendix A 
for highlighting the advantages of the $k$NN procedure. In addition, a functional real dataset is analyzed in Appendix B 
and a comparative study will show the interest of  semiparametrics. Technical proofs are gathered in Appendix C.
\section{The statistical methodology}
\subsection{A semiparametric functional regression model} \label{model}
Assume that we have a statistical sample of $n$ vectors $(X_{i1},\dots,X_{ip},\mathcal{X}_i,Y_i)$ ($i=1,\dots,n$) iid as $(X_1,\dots,X_p,\mathcal{X},Y)$, where $X_j$ ($j=1,\dots,p$) and $Y$ are real random variables, while $\mathcal{X}$ is a functional  random variable valued in a separable Hilbert space $\mathcal{H}$ with inner product denoted by $\left\langle \cdot, \cdot \right\rangle$.
There is need for building a model which takes care both of the functional predictor $\mathcal{X}$ by using single functional index ideas (see eg  \citealt{ait}, \citealt{chehm11} or \citealt{ma16}) and of the multivariate ones by using partial-linear ideas (see eg \citealt{anev06} or \citealt{fenx16}). Section 4.2i in \cite{linv18} provides wide sets of references for these models. The Semi-Functional Partial Linear Single-Index Model (SFPLSIM) aims to mix  functional and multivariate components, leading to the  relationship
\begin{equation}
	\label{eq_model}
	Y_i=X_{i1}\beta_{01}+\dots+X_{ip}\beta_{0p}+m\left(\left<\theta_0,\mathcal{X}_i\right>\right)+\varepsilon_i \ (i=1,\dots,n),
\end{equation}
where $\varepsilon_i$ is a random error verifying $\mathbb{E}\left(\varepsilon_i|X_{i1},\dots,X_{ip},\mathcal{X}_i\right)=0$. The vector $\pmb{\beta}_0=\left(\beta_{01},\dots,\beta_{0p}\right)^{\top}\in\mathbb{R}^p$, the functional direction $\theta_0\in \mathcal{H}$ and the  link real-valued function $m(\cdot)$ are supposed unknown.
To insure identifiability of model (\ref{eq_model}) we assume that $\left<\theta_0,\theta_0\right>=1$ and that, for some arbitrary $t_0$ in the domain of $\theta_0$, one has $\theta_0(t_0)>0$ (see eg \citealt{ait}; see also \citealt*{wang_2016} for other ways to insure identifiability).
\subsection{The $k$NN statistics}\label{estimates}
The $k$NN ideas have been used in early nonparametric one-dimensional literature to build location-adaptive smoothers (see eg \citealt{collomb} or \citealt{dev94}), and they have recently been extended for nonparametric  FDA (see eg \citealt{biacg10} and \citealt{kara} for recent results, and Section 2.2 in \citealt{linv18} for a survey). 
First of all, for each $\theta \in \mathcal{H}$, we define the operator $m_{\theta}(\cdot):\mathcal{H}\longrightarrow\mathbb{R}$ as $m_{\theta}(\chi)=m\left(\left<\theta,\chi\right>\right)$, $\forall\chi \in \mathcal{H}$. Note that $m_{\theta_0}(\mathcal{X})=\mathbb{E}\left(Y-\pmb{X}^{\top}\pmb{\beta}_0|\left<\theta_0,\mathcal{X}\right>\right)$, where $\pmb{X}=(X_{1},\dots,X_{p})^{\top}$. Then, $k$NN ideas are used for estimating $m_{\theta_0}(\cdot)$ from a nonzero smoothing factor $k=k_n\in \mathbb{N}$ and a kernel function $K$ as follows:
\begin{equation}
	\widehat{m}^\ast_{k,\theta,\pmb{\beta}}(\chi)=\sum_{i=1}^nw_{k,\theta}^*(\chi,\mathcal{X}_i)\left(Y_i-\pmb{X}_i^{\top}\pmb{\beta}\right)
	,\label{kNN_est}
\end{equation}
\noindent where, $\forall \chi\in\mathcal{H}$, we have denoted
$$w_{k,\theta}^*(\chi,\mathcal{X}_i)=\frac{K\left(H_{k,\chi,\theta}^{-1}d_{\theta}\left(\mathcal{X}_i,\chi\right)\right)}{\sum_{i=1}^nK\left(H_{k,\chi,\theta}^{-1}d_{\theta}\left(\mathcal{X}_i,\chi\right)\right)},
$$
with $d_{\theta}(\chi,\chi')=|\left<\theta,\chi-\chi'\right>| \ \forall \chi,\chi'\in\mathcal{H},$ and
$$
H_{k,\chi,\theta}=\min\left\{h\in \mathbb{R}^+ \mbox{\text{ such that }} \sum_{i=1}^n1_{B_\theta(\chi,h)}(\mathcal{X}_i)=k\right\}, 
$$
with $B_{\theta}(\chi,h)=\left\{z\in \mathcal{H}:d_{\theta}\left(\chi,z\right) \leq h\right\}$. It is worth being noted that this $k$NN statistic is an extension of the usual Nadaraya-Watson one,
\begin{equation}
	\widehat{m}_{h,\theta,\pmb{\beta}}(\chi)=\sum_{i=1}^nw_{h,\theta}(\chi,\mathcal{X}_i)\left(Y_i-\pmb{X}_i^{\top}\pmb{\beta}\right),  \label{kernel_est}
\end{equation}
where  $h\in \mathbb{R}^+$ is the bandwidth ($h=h_n$ depends on $n$) and 
$w_{h,\theta}(\chi,\mathcal{X}_i)=K\left(h^{-1}d_{\theta}(\mathcal{X}_i,\chi)\right)/\sum_{i=1}^nK\left(h^{-1}d_{\theta}(\mathcal{X}_i,\chi)\right)$. The $k$NN statistics presents, at least, two main advantages in practice in comparison with  the kernel one. On the one hand,
although the number of neighbours, $k$, is fixed, the bandwidth $H_{k,\chi,\theta}$ varies with $\chi$, providing the local-adaptive property of $k$NN-based estimators (allowing adaptation to heterogeneous designs). On the other hand, the selection of the smoothing parameter $k$ has lower computational cost than the selection of $h$, since $k$ takes values in the finite set $\{1,2,\dots,n\}$. However, the price to pay for these nice practical features is that, from a theoretical point of view, properties of the $k$NN statistics are much more difficult to obtain, mainly because  $H_{k,\chi,\theta}$ is a random variable depending on $\mathcal{X}_i$ ($i=1,\dots,n$) and avoiding for decomposing (\ref{kNN_est}) as sums of iid terms. Finally, it is worth being noted that, to estimate $m(\cdot)$ in (\ref{eq_model}) by means of (\ref{kNN_est}) and (\ref{kernel_est}), one needs to introduce in (\ref{kNN_est}) and (\ref{kernel_est}) estimates not only of $\theta_0$ (as in the case of the functional single-index model (FSIM); see \citealt{novo_2019}) but also of $\pmb{\beta}_0$. This fact is the major difficulty of the theoretical study of the estimator of $m(\cdot)$ presented in this paper compared to that of the FSIM.
\section{Some asymptotics} \label{theory}
\label{asymptotics}
\subsection{Technical assumptions}
In order to state results of uniform  (over $k$, $\theta$ and $\pmb{\beta}$) almost-complete consistency for  $\widehat{m}_{k,\theta,\pmb{\beta}}^*(\chi)$, the following technical assumptions will be needed:\\

\noindent-- We assume that the functional covariate is bounded in the following sense: 
\begin{equation}
	\exists C  \textrm{ such that }\left<\mathcal{X},\mathcal{X}\right>^{1/2}\leq C,\label{h_X_functional}
\end{equation}
(remember that $\left\langle \cdot, \cdot \right\rangle$ denotes the inner product associated to $\mathcal{H}$) and that the following condition on the conditional moments of the errors of the linear regression is verified:
\begin{eqnarray}
	\exists r\geq2, \ \exists C> 0 \textrm{ such that }
	\mathbb{E}\left(|Y-\pmb{X}^{\top}\pmb{\beta}_0|^r|\mathcal{X}\right)<C<\infty,\ a.s.\label{h_cond_mom_res}
\end{eqnarray}
Furthermore, let us denote by $N_{\chi,\theta_0}$ a fixed neighbourhood of $\chi\in\mathcal{H}$ in the topological space induced by the semi-metric $d_{\theta_0}(\cdot,\cdot)$, and denote $g_{j,\theta_0}(\chi)=\mathbb{E}\left(X_{ij}|\left\langle \theta_0, \mathcal{X}_i \right\rangle=\left\langle \theta_0,\chi\right\rangle\right)$ ($j=1,\dots,p$). 
H\"older type conditions are assumed for regression operators in the sense that exist constants $0\leq C<\infty$ and $\alpha_0>0$ such that, $\forall \chi_1,\chi_2\in N_{\chi,\theta_0},   \ \forall z\in\{m_{\theta_0},g_{1,\theta_0},\dots,g_{p,\theta_0}\}$,
\begin{eqnarray}
	\left|z(\chi_1)-z(\chi_2)\right|\leq Cd_{\theta_0}\left(\chi_1, \chi_2\right)^{\alpha_0}.\label{h_mg}
\end{eqnarray}
Furthermore, for fixed $\chi\in\mathcal{H}$ it is verified that
\begin{equation}
	\max_{j=1,\dots,p} \left|g_{j, \theta_0}(\chi)\right| =O\left(1 \right).\label{h_g2}
\end{equation}
-- It is assumed that the vector $\pmb{\beta}$ is not far from the target vector $\pmb{\beta}_0$, in the sense that there exists a sequence $\{c_n\}$, with $c_n\rightarrow0$ as $n\rightarrow\infty$, such that 
\begin{equation}
	\Psi_n=\left\{\pmb{\beta}\in\mathbb{R}^p; \lvert\lvert \pmb{\beta}-\pmb{\beta}_0\lvert\lvert=O(c_n)\right\}.\label{h_Phi_n}
\end{equation}
-- We assume that the cardinal of the space of directions, $\Theta_n$, verifies:
\begin{equation}
	\textrm{card}(\Theta_n)=n^{\alpha}\quad \textrm{with} \quad \alpha>0,
	\label{h_Theta_n}
\end{equation}
and that the elements of $\Theta_n$ are relatively close to the target direction $\theta_0$, in the sense that exists a sequence $\{b_n\}$ such that
\begin{equation}
	\forall \ \theta \in \Theta_n, \ \left<\theta-\theta_0,\theta-\theta_0\right>^{1/2} \leq C b_n.
	\label{h_theta}
\end{equation}
-- Let us define, for all $h>0$, $\theta\in\Theta_n$ and  $\chi\in\mathcal{H}$, the small ball probability function  $\phi_{\chi,\theta}(h)=\mathbb{P}\left(d_{\theta}(\mathcal{X}, \chi) \leq h\right)$
and assume that there exist constants $0<C_1\leq C_{2}<\infty$
and a function $f:\mathbb{R}\longrightarrow(0,\infty)$ such that 
\begin{equation}
	\forall \theta \in \Theta_n, \ C_1f(h)\leq\phi_{\chi,\theta}(h)\leq C_{2}f(h).
	\label{h_f1}
\end{equation}
Actually, it could be the case that $f(\cdot)=f_{\chi}(\cdot)$ (for sake of brevity we omit the sub-index $\chi$). To control the variance of the estimators, it is assumed that   there exist constants $0<C_{1}\leq C_{2}<\infty$ and sequences $\{a_n\},\{b_n\} \subset \mathbb{R}^+$ ($a_n \leq b_n$) such that, for $h \in [a_n,b_n]$ 
\begin{equation}
	C_{1} \leq \frac{f(h/2)}{f(h)}\leq C_{2}, {\mbox{ for $n$ large enough}}.
	\label{h_f2}
\end{equation}
It is assumed that there exist sequences $\{\rho_n\} \subset (0,1)$, $\{k_{1,n}\}\subset \mathbb{Z}^+$, $\{k_{2,n}\} \subset \mathbb{Z}^+$ ($k_{1,n}\leq k_{2,n}\leq n$) and constants $0<\lambda\leq \delta<\infty$ satisfying:
\begin{equation}
	\lambda f^{-1}\left(\frac{\rho _n k_{1,n}}{n}\right)\leq\phi_{\chi,\theta}^{-1}\left(\frac{\rho_n k_{1,n}}{n}\right) \mbox{\text{ and }}
	\phi_{\chi,\theta}^{-1}\left(\frac{ k_{2,n}}{\rho_n n}\right) \leq \delta f^{-1}\left(\frac{ k_{2,n}}{\rho_n n}\right),
	\label{h_k1}
\end{equation}
\begin{equation}
	f^{-1}\left(\frac{ k_{2,n}}{\rho_n n}\right) \rightarrow 0, \
	\min \left\{\frac{1-\rho_n}{4}\frac{k_{1,n}}{\ln n},\frac{(1-\rho_n)^2}{4\rho_n}\frac{k_{1,n}}{\ln n}\right\} > \alpha +2, \label{h_k3}
\end{equation}	
\begin{equation}
	\frac{\log n}{n \min\left\{\lambda f^{-1}(\rho_n k_{1,n}/n), f\left(\lambda f^{-1}(\rho_n k_{1,n}/n)\right) \right\}}\rightarrow 0 \label{h_k4}.
\end{equation}
--We assume that the  kernel function, $K$, verifies
\begin{equation}
	0<C_1 1_{(0,1/2)}(\cdot) \leq K(\cdot) \leq C_2 1_{(0,1/2)}(\cdot),\label{h_K}
\end{equation}
where $1_{(0,1/2)}$ denotes the indicator function of the set $(0,1/2)$. In addition, one controls the complexity of the following  classes of functions:
\begin{equation*}
	\mathcal{K}_{\theta} = \left\{\cdot\longrightarrow K\left(h^{-1}d_{\theta}(\chi,\cdot)\right), \ h > 0\right\},\end{equation*}
by assuming that  $\mathcal{K}_{\Theta_n} = \cup_{\theta \in \Theta_n}\mathcal{K}_{\theta}$ is a pointwise measurable class such that
\begin{equation}
	\sup_{\mathcal{Q}}\int_{0}^1\sqrt{1+\log\mathcal{N}\left(\epsilon\lVert F_{\Theta_n}\lVert_{\mathcal{Q},2},\mathcal{K}_{\Theta_n},d_{\mathcal{Q},2}\right)}d\epsilon<\infty.
	\label{h_N_K}
\end{equation}
Note that in (\ref{h_N_K}), $F_{\Theta_n}$ is the minimal envelope function of the set $\mathcal{K}_{\Theta_n}$, the supremum is taken  over all probability measures ($\mathcal{Q}$) on the measurable  space  $(\mathcal{H},\mathcal{A})$ with $||F_{\Theta_n}||_{\mathcal{Q},2}^2<\infty$, $||\cdot||_{\mathcal{Q},2}$ is the norm $L_2(\mathcal{Q})$ defined on $S=\{f:\mathcal{H}\longrightarrow \mathbb{R}\}$, and $d_{\mathcal{Q},2}(\cdot,\cdot)$ is the metric associated to the norm $L_{2}(\mathcal{Q})$.
Finally, given a metric space $(\mathcal{K},d)$, $\mathcal{N}\left(\epsilon,\mathcal{K},d\right)$ denotes the minimal number of open balls, in the topological space given by $d$, with radius $\epsilon$ which are needed to cover $\mathcal{K}$ (see \citealt{novo_2019} for details).

The large number of hypotheses, allowing to deal with the complexity of the model and to obtain general results, are actually not very restrictive. On one hand, (\ref{h_cond_mom_res}), (\ref{h_mg}), (\ref{h_g2}) and   (\ref{h_K}) are standard assumptions in regression models mixing linear and nonparametric structures (see eg \citealt{anev06}). On the other hand, 
(\ref{h_X_functional}), (\ref{h_Theta_n})-(\ref{h_k4}) and (\ref{h_N_K}) are assumptions being usual for obtaining uniform consistency of any $k$NN-based estimators (see \citealt{kara} in nonparametric or \citealt{novo_2019} in semiparametric models). Finally, Assumption (\ref{h_Phi_n}) is added for controlling the bias in the estimation of the linear coefficients in model (\ref{eq_model}).
\subsection{Uniform rates of consistency of $k$NN estimates.}
\label{uni-rat}
The next Theorem \ref{theorem} is the main part of this paper. 
\begin{theorem}
	\label{theorem}
	Under conditions (\ref{eq_model}) and (\ref{h_X_functional})- (\ref{h_N_K}), we have that
	\begin{eqnarray}
		\sup_{\pmb{\beta}\in \Psi_n}\sup_{\theta\in\Theta_n}\sup_{k_{1,n}\leq k\leq k_{2,n}}|\widehat{m}^\ast_{k,\theta,\pmb{\beta}}(\chi)-m_{\theta_0}(\chi)|=O\left(f^{-1}\left(\frac{k_{2,n}}{\rho_n n}\right)^{\alpha_0}\right)\nonumber\\+O_{a.co.}\left(\sqrt{\frac{\log n}{nf\left(\lambda f^{-1}(\rho_n k_{1,n}/n)\right)}}\right)+O\left(c_n\right).\nonumber
	\end{eqnarray}	
\end{theorem}
Note that the first two terms in these rates of convergence are the same as those in \cite{novo_2019} in the simpler model without multivariate predictors. The third term in the rates corresponds to the bias when estimating the linear coefficients of the model. For small enough values of $c_n$ this third term could be much smaller than both previous ones, highlighting the fact that the presence of linear component in the SFPLSIM does not deteriorate the asymptotics. Note also that, under standard additional conditions on $f(\cdot),\ \rho_n,\ k_{1,n}$ and $k_{2,n}$ (see eg \citealt{novo_2019}), the rates in Theorem \ref{theorem} are the same as if  $\mathcal{X}$ was one-dimensional: with other words,  the semiparametric model has achieved its goal of being unsensitive to dimensionality effects. 

Theorem \ref{theorem} is confirming the well-known fact that practical using of the method is linked with a  choice of the smoothing factor $k$  balancing the trade-off between bias and variance effects. One of the most important feature of our result is to be uniform over $k \in [k_{1,n}k_{2,n}]$, $\pmb{\beta}\in \Psi_n$ and $\theta\in\Theta_n$, allowing to say that the same asymptotics are available when $k,\ \pmb{\beta}$ and $\theta$ are random variables valued in $[k_{1,n}k_{2,n}], \Psi_n$ and $\Theta_n$, respectively (in particular when $k, \ \pmb{\beta}$ and $\theta$ are data-driven selected). This is formulated in the next corollary, whose proof is obvious (because of the uniform feature of previous theorem), making the proposed methodology fully automatic in the sense that the main parameter ($k$) as well as the two other ones ($\pmb{\beta}$ and $\theta$) can be selected from the sample without deteriorating its asymptotic behaviour.
\begin{corollary}
	\label{cor}
	Assume that the conditions of Theorem  \ref{theorem} hold. Assume that $\hat{k}, \ \hat{\pmb{\beta}}$ and $\hat{\theta}$ are random variables taking values in $[k_{1,n}k_{2,n}], \ \Psi_n$ and $\Theta_n$, respectively, being data-driven in the sense that they depend on the statistical sample $S_n=\{(X_{i1},\dots,X_{ip},\mathcal{X}_i,Y_i), \ i=1,\ldots,n\}$ (i.e. $\hat{k}=\hat{k}(S_n), \ \hat{\pmb{\beta}}=\hat{\pmb{\beta}}(S_n)$ and $\hat{\theta}=\hat{\theta}(S_n)$). Then we have:
	\begin{eqnarray}
		|\widehat{m}^\ast_{\hat{k},\hat{\theta},\hat{\pmb{\beta}}}(\chi)-m_{\theta_0}(\chi)|=O\left(f^{-1}\left(\frac{k_{2,n}}{\rho_n n}\right)^{\alpha_0}\right)\\+O_{a.co.}\left(\sqrt{\frac{\log n}{nf\left(\lambda f^{-1}(\rho_n k_{1,n}/n)\right)}}\right)+O\left(c_n\right).\nonumber
	\end{eqnarray}
\end{corollary}
This corollary allows to have asymptotics for any automatic data-driven parameters. To fix the ideas let us just mention one example. Estimators $\hat{\theta}_k$ and $\hat{\pmb{\beta}_k}$ could be constructed from the ordinary least squares (OLS) procedure applied to a linear model in which the effects of the functional covariate have been extracted. That is, $\hat{\theta}_k$ and $\hat{\pmb{\beta}_k}$ are minimizing the score function
\begin{equation}
	\mathcal{Q}^*_k\left(\pmb{\beta},\theta\right)=\frac{1}{2}\left(\widetilde{\pmb{Y}}^*_{k,\theta}-\widetilde{\pmb{X}}^*_{k,\theta}\pmb{\beta}\right)^{\top}\left(\widetilde{\pmb{Y}}^*_{k,\theta}-\widetilde{\pmb{X}}^*_{k,\theta}\pmb{\beta}\right),
	\label{func_minimizar}
\end{equation}
where $\pmb{X}=\left(\pmb{X}_1,\dots,\pmb{X}_n\right)^{\top}$, with $\pmb{X}_i=\left(X_{i1},\ldots,X_{ip}\right)^{\top}$, and $\pmb{Y}=\left(Y_1,\dots,Y_n\right)^{\top}$, while for any $(n\times q)$-matrix $\pmb{A}$ $(q\geq 1)$, number of neighbours $k$ and $\theta\in\Theta_n$, we denote
$\widetilde{\pmb{A}}_{k,\theta}^*=\left(\pmb{I}-\pmb{W}^*_{k,\theta}\right)\pmb{A}, \mbox{ where } \pmb{W}^*_{k,\theta}=\left(w^*_{k,\theta}(\mathcal{X}_i,\mathcal{X}_j)\right)_{i,j}$. 
Then cross-validation ideas (either leave-one-out or k-fold cross-validation) could be used to obtain an estimate $\hat{k}$ (for more specific details to put in practice our method, see Appendix A.1.

\subsection{Some by-product on asymptotics for kernel type estimates}

Even if our main purpose is to study $k$NN estimate, note that the same results can be obtained for the kernel estimate (\ref{kernel_est}) by changing (\ref{h_k1})-(\ref{h_k4}) into assuming that $\{a_n\}$ and $\{b_n\}$ are satisfying
\begin{equation}
	a_n\rightarrow 0 , b_n\rightarrow 0 \mbox{\text{ and }}  \frac{\log n}{n\min\left\{a_n,f(a_n)\right\}}\rightarrow 0 \label{h_a_b}.
\end{equation} 
The next Theorem \ref{theorem2} follows by proofs similar to those of Theorem \ref{theorem}. 
\begin{theorem}
	\label{theorem2}
	If the conditions of Theorem \ref{theorem} hold when changing (\ref{h_k1})-(\ref{h_k4}) into (\ref{h_a_b}), then
	\begin{equation*}
		\sup_{\pmb{\beta}\in \Psi_n}\sup_{\theta\in\Theta_n}\sup_{a_n\leq h\leq b_n}|\widehat{m}_{h,\theta,\pmb{\beta}}(\chi)-m_{\theta_0}(\chi)|=O\left(b_n^{\alpha_0}\right)+O_{a.co.}\left(\sqrt{\frac{\log n}{nf(a_n)}}\right)+O\left(c_n\right).
	\end{equation*}
\end{theorem}

\noindent {\bf{Acknowledgements}}

\noindent Authors wish to thank an Associate Editor and a Reviewer for helpful suggestions. 
This research was supported by MINECO grant MTM2017-82724-R and by the Xunta de Galicia (Grupos de Referencia Competitiva ED431C-2020-14 and Centro de Investigación del Sistema Universitario de Galicia ED431G 2019/01), all of them through the ERDF.

\newpage

\appendix

\begin{center}
{\large{\textbf{A $k$NN procedure in semiparametric functional data analysis}}}

Silvia Novo$^a$\footnote{Corresponding author email address: \href{s.novo@udc.es}{s.novo@udc.es} } \hspace{2pt} Germ\'an Aneiros$^b$ \hspace{2pt} Philippe Vieu$^c$
\end{center}

\noindent$^a$ Department of Mathematics, MODES, CITIC, Universidade da Coruña, A Coruña, Spain \\
\noindent$^b$ Department of Mathematics, MODES, CITIC, ITMATI, Universidade da Coruña, A Coruña, Spain\\
\noindent$^c$ Institut de Math\'{e}matiques, Universit\'e  Paul Sabatier,  Toulouse, France		
	
\bigskip

\bigskip

\begin{center}
\textbf{SUPPLEMENTARY MATERIAL} 
\end{center}

These pages contain a simulation study, an application to real data and the proofs of the asymptotic results presented in our paper. The used notation, as well as assumptions, enunciates of the theorems and the references, can be found in the paper. All the enumeration corresponding to the paper is maintained here (this includes enumeration related to equations, assumptions and theorems).

\section{Simulation study}\label{sim}
\subsection{The design}
\label{desi}
Samples of iid data  $\mathcal{D}=\{(X_{i1},X_{i2},X_{i3},\mathcal{X}_i,Y_i)\}_{i=1}^{n+25}$ were generated from the model
\begin{equation}
\label{eq_model_sim}
Y_i=X_{i1}\beta_{01}+X_{i2}\beta_{02}+X_{i3}\beta_{03}+\alpha m\left(\left<\theta_0,\mathcal{X}_i\right>\right)+(1-\alpha)r(\mathcal{X}_i)+\varepsilon_i.
\end{equation}
(Note that the case $\alpha=1$ gives the SFPLSIM studied in this paper, while values $\alpha \in [0,1)$ allow to show a sensitivity analysis of the proposed method.) The functional covariate, $\mathcal{X}_i$ ($i=1,\ldots,n+25$), was generated in the following way:
$\mathcal{X}_i(t)=a_i\cos(2\pi t) + b_i\sin(4\pi t) + 2c_i(t-0.25)(t-0.5) \ \forall t \in [0,1]$. To build heterogeneous curves dataset, the random variables $a_i,b_i$ and $c_i$ were independent variables being uniformly distributed either on $[5,10]$ with probability $0.5$ or on $[20,20.5]$ with probability $0.5$ (note that  independence means both between and within vectors $(a_i,b_i,c_i)^\top$). These curves were discretized on the same grid of $100$ equispaced points in $[0,1]$.  On the other hand, the vector of real covariates, $(X_{i1},X_{i2},X_{i3})^{\top}$ ($i=1,\ldots,n+25$), were generated from a multivariate normal distribution with zero mean and covariance matrix given by $(\rho^{|j-k|})_{jk}$ ($j,k=1,2,3$). 
The iid random errors, $\varepsilon_i$ ($i=1,\ldots,n+25$), were simulated from a $N(0,\sigma_{\varepsilon}^2=c\sigma_r^2)$ where $\sigma_r^2$ is the empirical variance of the regression function in (\ref{eq_model_sim}). The signal-to-noise ratio $c$ has been taken equal to $c=0.025$.

The true vector of linear coefficients was $\pmb{\beta}_0=(\beta_{01},\beta_{02},\beta_{03})^{\top}=(-1,0.5,1.5)^{\top},$ 
while the true direction of projection was \vspace{-0.3cm} 
\begin{equation}
\theta_0(\cdot)=\sum_{j=1}^{d_n}\alpha_{0j}e_j(\cdot), 
\label{theta_0_base}
\end{equation} 
where $\{e_1(\cdot),\ldots,e_{d_n}(\cdot)\}$ is a set of B-spline basis functions and $d_n=l+m_n$ ($l$ denotes the order of the splines and $m_n$ is the number of regularly interior knots). Values $l=3$ and $m_n=3$ were considered and the vector of coefficients of $\theta_0$ in expression (\ref{theta_0_base}) was obtained by calibrating the vector $(1,1,1,1,0,0)^\top$ in order to insure identifiability, and was equal to $(\alpha_{01},\dots,\alpha_{0d_n})^\top=(1.201061, 1.201061, 1.201061, 1.201061, 0, 0)^\top$.
Finally, $\left<f,g\right>=\int_{0}^1 f(t)g(t)dt$, $m(\left<\theta_0,\chi_i\right>)=\left<\theta_0,\chi_i\right>^3$ and $r(\chi_i)=2\sqrt{c_i}$ were considered (note that $\mathcal{X}_i=\mathcal{X}_{a_i,b_i,c_i}$). 

For each simulation case $(n,\rho,\alpha)\in\{50,100,200\}\times\{0,0.5\}\times\{0.8,0.9,1\}$, $M=100$ independent samples were generated from (\ref{eq_model_sim}). Each sample ${\cal{D}}$ was split into two subsamples: a training sample, ${\cal{D}}_{n,train}=\{(X_{i1},X_{i2},X_{i3},\mathcal{X}_i,Y_i)\}_{i=1}^{n}$, and a testing sample, ${\cal{D}}_{n,test}=\{(X_{i1},X_{i2},X_{i3},\mathcal{X}_i,Y_i)\}_{i=n+1}^{n+25}$. The tuning parameters ($\hat{h}$ and $\hat{k}$)  were constructed from the training sample by means of the 10-fold cross-validation procedure. In addition, we only use  the training sample for getting estimations of $\theta_0$ ($\widehat{\theta}_0$ with the kernel-based method and $\widehat{\theta}_0^\ast$ with the $k$NN-based one) and of $\pmb{\beta}_{0}$ ($\widehat{\pmb{\beta}}_0$ with the kernel-based procedure and $\widehat{\pmb{\beta}}_0^\ast$ with the $k$NN-based one). These $k$NN-based estimations were obtained by minimizing the score function (\ref{func_minimizar}), as suggested at the end of Section \ref{uni-rat}; the same procedure was used to construct the kernel-based ones, introducing the obvious modifications in (\ref{func_minimizar}) ($k$ and $w_{k,\theta}^*(\cdot,\cdot)$ should be replaced by $h$ and $w_{h,\theta}(\cdot,\cdot)$, respectively). For constructing in practice the set of eligible directions $\Theta_n$, we considered as eligible functional directions
$\theta(\cdot)=\sum_{j=1}^{d_n}\alpha_je_j(\cdot)$
for a wide set of vectors of coefficients, $(\alpha_1,\dots,\alpha_{d_n})^\top$, constructed  following the procedure described in \cite{novo_2019}.

For measuring the performance of the proposed estimators we computed \vspace{-0.2cm} 
\begin{equation*}
||\widehat{\pmb{\beta}}_0-\pmb{\beta}_0||^2=\sum_{j=1}^{3}(\widehat{\beta}_{0j} - \beta_{0j})^2,\quad
||\widehat{\pmb{\beta}}_0^*-\pmb{\beta}_0||^2=\sum_{j=1}^{3}(\widehat{\beta}_{0j}^* - \beta_{0j})^2,
\label{error-beta0}
\end{equation*} 
\begin{equation*}
||\widehat{\theta}_0-\theta_0||^2=\int_0^1\left(\widehat{\theta}_{0}(t)-\theta_0(t)\right)^2dt,\quad
||\widehat{\theta}_0^*-\theta_0||^2=\int_0^1\left(\widehat{\theta}_{0}^*(t)-\theta_0(t)\right)^2dt,
\label{error-theta0}
\end{equation*}

\begin{equation}
\label{MSEP-sim}
\textrm{ and } \ MSEP_n=\frac{1}{n_{test}}\sum_{i=n+1}^{n+n_{test}}(Y_i -\widehat{Y}_i)^2,
\end{equation}
where $\widehat{Y}_i$ denotes a predicted value for $Y_i$ (here one has $n_{test}=25$).
\subsection{Results}
The results are summarized in Tables \ref{table-MSEP}, \ref{table-error-beta} and \ref{table-error-theta} below. On the one hand, it appears that both methods are benefited by the increase of the sample size. More importantly it seems that, as well for independent covariates ($\rho=0$)  as for correlated  ones ($\rho=0.5$), the $k$NN-based procedure clearly overpasses results obtained with the kernel-based procedure by being able to capture heterogeneous structure of the data. Finally, the proposed procedure is not very sensitive, at least in this example, to slight modifications (high values of $\alpha$) in the effect of the functional covariate.

\begin{table}[H]
\caption{Averaged MSEPs with 
	10-fold cross-validation selectors for $h$ and $k$}
\label{table-MSEP}
\centering
\scalebox{0.85}{
	\begin{tabular}{cccclcclcc}
		\hline
		&&\multicolumn{2}{c}{$n=50$} &  & \multicolumn{2}{c}{$n=100$} &  & 
		\multicolumn{2}{c}{$n=200$} \\ \cline{3-4}  \cline{6-7}  \cline{9-10}
		&	&kernel & $k$NN &  & kernel & $k$NN &  & kernel & $k$NN \\ 
		\hline
		\multirow{2}{*}{$\alpha$=1}&	$\rho=0$ &	0.1959 &  0.1626&  & 0.1619 &0.1297  &  &  0.1239 &  0.1024 \\
		&	$\rho=0.5$ & 0.1791 &0.1393 & &0.1458 &0.1154 & & 0.1068 & 0.0893\\
		\hline
		\multirow{2}{*}{$\alpha$=0.9} &	$\rho=0$ & 0.2088  
		&  0.1785 & &0.1674 &0.1431 & & 0.1350&0.1121\\ &$\rho=0.5$ & 0.1838& 0.1583  & &  0.1500 &0.1278  & &  0.1187&  0.0992 \\
		\hline
		\multirow{2}{*}{$\alpha$=0.8}&	$\rho=0$ & 0.2193 &0.1976 & &0.1858 &0.1591 & &0.1473 &  0.1200 \\ 	&$\rho=0.5$& 0.2016 & 0.1767 &  & 0.1654 &0.1426 & &0.1307 & 0.1067 \\
		\hline
\end{tabular}}
\end{table}

\begin{table}[H]
\caption{Averaged squared errors for $\pmb{\beta}_0$}
\label{table-error-beta}
\centering
\scalebox{0.85}{\begin{tabular}{cccclcclcc}
		\hline
		&&\multicolumn{2}{c}{$n=50$} &  & \multicolumn{2}{c}{$n=100$} &  & 
		
		\multicolumn{2}{c}{$n=200$} \\ 
		\cline{3-4}  \cline{6-7}  \cline{9-10}
		
		&	&kernel & $k$NN &  & kernel & $k$NN &  & kernel & $k$NN \\ 
		\hline
		\multirow{2}{*}{$\alpha$=1} &	$\rho=0$ &0.0133 & 0.0097&  & 0.0043 & 0.0041  &  &  0.0021&  0.0018 \\
		&	$\rho=0.5$ & 0.0181 & 0.0120 & &0.0059&0.0058 & & 0.0025 & 0.0021  \\     
		
		\hline
		\multirow{2}{*}{$\alpha$=0.9} &	$\rho=0$ &0.0140 
		& 0.0105 & & 0.0047 &0.0044 & & 0.0022 & 0.0020 \\ &$\rho=0.5$ & 0.0183 & 0.0138 & &0.0063 & 0.0064 & & 0.0026& 0.0024\\
		\hline
		\multirow{2}{*}{$\alpha$=0.8}&	$\rho=0$ & 0.0141 & 0.0117 & &0.0049 & 0.0047 & &   0.0025&  0.0022\\ 	&$\rho=0.5$& 0.0187&  0.0154 & &0.0067   &0.0069 & & 0.0029& 0.0028\\
		\hline
\end{tabular}}
\end{table}

\begin{table}[H]
\caption{Averaged squared errors for $\theta_0$}
\label{table-error-theta}
\centering
\scalebox{0.85}{\begin{tabular}{cccclcclcc}
		\hline
		&	&\multicolumn{2}{c}{$n=50$} &  & \multicolumn{2}{c}{$n=100$} &  & 
		\multicolumn{2}{c}{$n=200$} \\	\cline{3-4}  \cline{6-7}  \cline{9-10}
		
		&	&kernel & $k$NN &  & kernel & $k$NN &  & kernel & $k$NN \\ 
		\hline
		\multirow{2}{*}{$\alpha$=1} &	$\rho=0$ &0.0950 & 0.0507&  & 0.0715 & 0.0413 &  &  0.0603& 0.0070 \\
		& $\rho=0.5$ &0.0933&  0.0463 & &0.0659&0.0389& &  0.0618 & 0.0061  \\
		\hline
		\multirow{2}{*}{$\alpha$=0.9} &	$\rho=0$ & 0.0958
		& 0.0656 & & 0.0713 & 0.0595& &0.0679  &0.0330\\       &$\rho=0.5$ & 0.0931 &  0.0622  & & 0.0697 &0.0586 & &0.0643 &0.0302\\
		\hline
		\multirow{2}{*}{$\alpha$=0.8}&	$\rho=0$ &  0.0921 & 0.0781 & & 0.0871 &0.0759 & &   0.0732&0.0757\\ 	&$\rho=0.5$& 0.0895 & 0.0758  & & 0.0851&0.0746 & &  0.0756 & 0.0751\\
		\hline
\end{tabular}}
\end{table}
\vspace{-0.3cm}
\vspace{-0.3cm}
\section{Real data application}\label{appli}
\begin{figure}[H]
\centering
\includegraphics[width=0.30\textwidth]{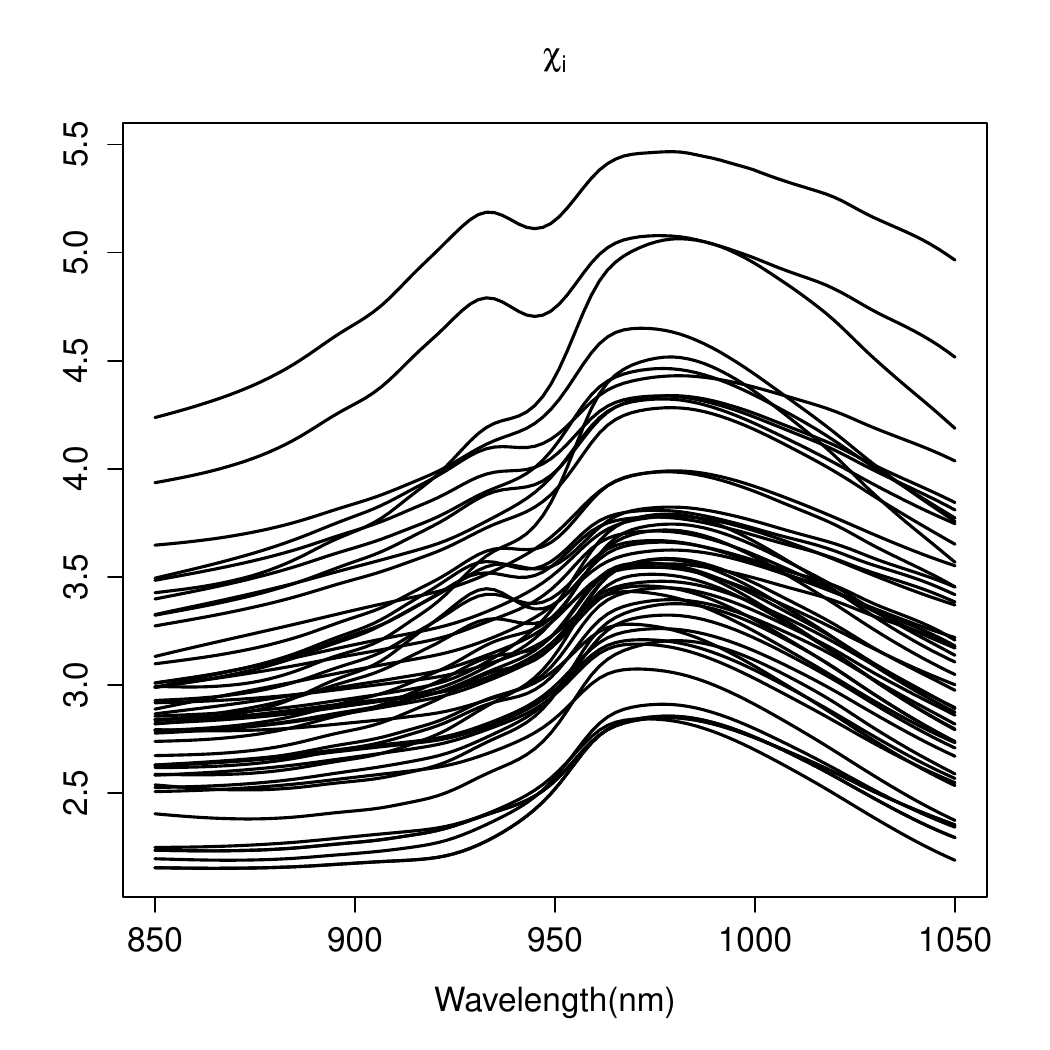}\hspace{1.5cm}
\includegraphics[width=0.30\textwidth]{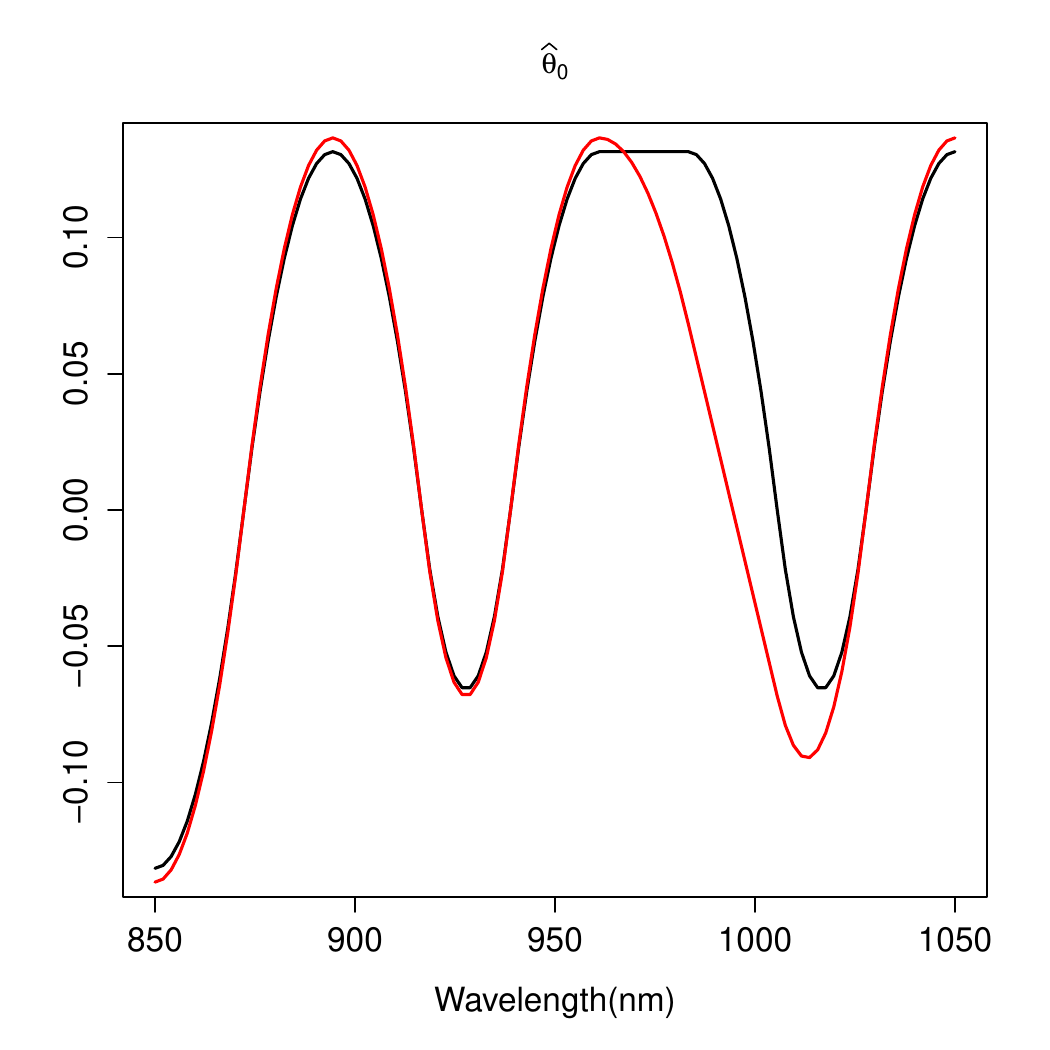}\vspace{-0.50cm}
\caption{Left panel: Sample of 50 absorbance curves $\mathcal{X}_i$. Right panel: Estimates of the functional direction $\theta_0$ using $k$NN-based (red line) and kernel-based (black line) estimators.}
\label{fig5}
\end{figure}

This section is devoted to illustrate the usefulness of the SFPLSIM (\ref{eq_model}), as well as to compare the performance of kernel and $k$NN procedures. We will analyse the benchmark ``Tecator's data", which contains measurements of contents of fatness ($Y_i$), of protein ($X_{1i}$) and of  moisture  ($X_{2i}$)  for $215$  pieces of meat as well as  the near-infrared absorbance spectras ($\mathcal{X}_i$) observed on 100 equally wavelengths in the range $850-1050$ nm. The left panel in Figure \ref{fig5} shows a sample of 50 absorbance curves. 
Our purpose is to model the link between fat content and the other variables, with aim to predict the fat content. We will split the original sample into two subsamples:
a training sample, $\mathcal{D}_{train}=\{(X_{i1},X_{i2},\mathcal{X}_i,Y_i)\}_{i=1}^{160},$ and a testing one, $\mathcal{D}_{test}=\{(X_{i1},X_{i2},\mathcal{X}_i,Y_i)\}_{i=161}^{215}$. The estimation task is made only by means of the training sample, while the testing sample is used to measure the quality of the predictions. So, to quantify the prediction error  we use the MSEP (see (\ref{MSEP-sim})) with $n_{test}=55$.

Firstly, we predict the fat content of meat using two simple models involving only the two scalar covariates: a bivariate linear model (LM) and an additive spline model (ASM). Both models give similar results which are reported 
in Table \ref{table-mod-scalar}.

\begin{table}[H]
\caption{ MSEP  for models with two scalar covariates.}
\label{table-mod-scalar}
\centering
\scalebox{0.85}{\begin{tabular}{cccccc}
		\hline
		&&\multicolumn{2}{c}{Model} & &  MSEP \\
		\hline
		LM: && $Y=\beta_{01}X_1+\beta_{02}X_2+\varepsilon$ & & &1.95\\
		ASM: && $Y=r(X_1)+r(X_2)+\varepsilon$ & & &1.93\\
		\hline
\end{tabular}}	
\end{table}
\vspace{-0.2cm}
\noindent In addition, we report in Table  \ref{table-mod-functional} the results obtained with simple models involving only the functional covariate, such as the functional linear model (FLM), the functional nonparametric model (FNM), the FSIM, and  the FSIM combined with the application of a full nonparametric boosting step to its residuals (FSIM \& FNM, for details see \citealt{novo_2019}).
One observes that  $k$NN-based estimation overpasses kernel-based one in each case, but with $k$NN each model gives results being more or less similar to those of models in Table \ref{table-mod-scalar}.

\begin{table}[H]
\caption{ Values of the MSEPs for some functional models.}
\label{table-mod-functional}
\centering
\scalebox{0.85}{\begin{tabular}{llll}
		\hline
		& Model & \multicolumn{2}{c}{MSEP} \\ \hline
		
		FLM: & $Y=\alpha _{0}+\int_{850}^{1050}\mathcal{X}^{(2)}(t)\alpha (t)dt+\varepsilon $
		& \multicolumn{2}{c}{7.17} \\ \hline
		&  & kernel & $k$NN \\ \cline{3-4}
		FNM: & $Y=r(\mathcal{X}^{(2)})+\varepsilon $ & $4.06$ & $1.79$ \\ 
		
		FSIM: & $Y=m\left( \left\langle \theta _{0},\mathcal{X}^{(2)}\right\rangle \right)
		+\varepsilon $ & $3.49$ & $2.69$ \\ 
		
		FSIM \& FNM (boosting step):  & $Y=m\left( \left\langle \theta _{0},\mathcal{X}^{(2)}\right\rangle \right)+r(\mathcal{X}^{(1)})
		+\varepsilon $ & $1.74$ & $1.53$ \\ 
		\hline
\end{tabular}}
\end{table}
\vspace{-0.2cm}

\noindent Finally, we used models incorporating both scalar and functional  covariates, namely  the SFPLM and the SFPLSIM (\ref{eq_model}) proposed in this paper. For both models, we use OLS-based estimators for estimating $\pmb{\beta}_0$ (and also $\theta_0$ in the SFPLSIM case) and 10-fold cross-validation  for selecting $k$, $h$, the order $q$ of the derivatives of the absorbance curves ($\mathcal{X}_i^{(q)}$) and the number $m_n$ of regularly interior knots of the B-spline basis functions considered to construct the set of eligible directions $\Theta_n$ (for details, see Section \ref{desi}). Table \ref{table-mod-funct-semi} is summarizing the results. In both cases, the $k$NN-based estimation procedures overpass the kernel-based ones and the SFPLSIM offers lower MSEP than the SFPLM. More importantly, these models involving both kinds of covariate gives much smaller prediction error that models using only one kind of variables (as those in Tables \ref{table-mod-scalar} and \ref{table-mod-functional}). All in all, the SFPLSIM model with $k$NN estimates leads to the lowest MSEP among all models/estimates studied.

\begin{table}[H]
\caption{Values of the MSEPs for some  functional partial linear models}
\label{table-mod-funct-semi}
\centering
\scalebox{0.85}{\begin{tabular}{cll}
		\hline
		Model & \multicolumn{2}{c}{MSEP} \\ \hline
		&kernel & $k$NN  \\ \cline{2-3}
		SFPLM:  $Y=\beta_{01}X_1+\beta_{02}X_2+r(\mathcal{X}^{(1)})+\varepsilon $ & 0.87& 0.69 \\ 
		SFPLSIM:  $Y=\beta_{01}X_1+\beta_{02}X_2+m\left( \left\langle \theta _{0},\mathcal{X}^{(1)}\right\rangle \right)
		+\varepsilon $ & 0.77 & 0.60 \\ 
		\hline
\end{tabular}}
\end{table}

To conclude, it is worth being pointed that in addition to this good predictive behaviour, another great advantage of the SFPLSIM  is that the functional variable enters in the model through interpretable parameter: $\theta_0$. The obtained estimations of this functional direction in the SFPLSIM, using both $k$NN and kernel-based estimation procedures, can be seen in Figure \ref{fig5} (right panel). The estimated directions show two peaks and two bumps which could give information on which wavelengths ranges have the highest influence on the fat content. We also would like to remember that, to obtain our estimate of $\theta_0$, our method proposes to minimize on a predefined index set $\Theta_n$. Therefore, its computational cost is higher than the one of efficient proposals based on the use of functional dimension reduction techniques, as that in \citealt{wang_2016}. The advantage of our method against such proposals is (at least in this example) its great predictive power: considering the same both Tecator subsamples and measure of the predictive performance as in \citealt{wang_2016}, our procedure improves in a $35\%$ the predictive power of the method in \citealt{wang_2016}.

\section{Proofs}\label{app}

\subsection{Proof of Theorem \ref{theorem}}
Let us first introduce some additional notation.  
The $k$NN statistics associated with the estimation of $g_{j,\theta_0}(\cdot)$ $(j=1,\dots,p)$, for each $\theta\in\Theta_n$ will be defined as:
\begin{eqnarray}
\widehat{g}^*_{j,k,\theta}(\chi)=\sum_{i=1}^n w_{k,\theta}^*(\chi,\mathcal{X}_i)X_{ij} \ \forall \chi \in \mathcal{H}.\nonumber
\end{eqnarray}
The main idea of the proof consists in applying existing results for $k$NN estimates in the single functional index model  without additional multivariate predictors, and then to deal with the question of estimating the additional linear coefficients $\beta$. For fixed $\chi\in\mathcal{H}$, the following decomposition can be made:
\begin{eqnarray}
\left|\widehat{m}_{k,\theta,\pmb{\beta}}^*(\chi)-m_{\theta_0}(\chi)\right|
\leq \left|\sum_{i=1}^n w_{k,\theta}^*(\chi,\mathcal{X}_i)\left(m_{\theta_0}(\mathcal{X}_i)+\varepsilon_i\right)-m_{\theta_0}(\chi)\right| \label{PKNN1} \\+\left|\sum_{j=1}^pg_{j,\theta_0}(\chi)\left(\beta_{0j}-\beta_j\right)\right|+\left|\sum_{j=1}^p \left(\widehat{g}_{j,k,\theta}^*(\chi)-g_{j,\theta_0}(\chi)\right)\left(\beta_{0j}-\beta_j\right)\right|.\nonumber
\end{eqnarray}

Now, using Theorem 3.3(b) in \cite{novo_2019}, it is obtained that
\begin{eqnarray}
\sup_{\theta\in\Theta_n}\sup_{k_{1,n}\leq h\leq k_{2,n}}\left|\sum_{i=1}^n w_{k,\theta}^*(\chi,\mathcal{X}_i)\left(m_{\theta_0}(\mathcal{X}_i)+\varepsilon_i\right)-m_{\theta_0}(\chi)\right|=\nonumber\\O_{a.co.}\left(\sqrt{\frac{\log n}{nf\left(\lambda f^{-1}(\rho_n k_{1,n}/n)\right)}}\right)+O\left(f^{-1}\left(\frac{k_{2,n}}{\rho_n n}\right)^{\alpha_0}\right).\label{PKNN2}
\end{eqnarray}

\noindent Now, using again Theorem 3.3(b) in \cite{novo_2019} together with condition (\ref{h_Phi_n}), one has
\begin{eqnarray}
\left|\sum_{j=1}^p \left(\widehat{g}_{k,j,\theta}^*(\chi)-g_{j,\theta_0}(\chi)\right)\left(\beta_{0j}-\beta_j\right)\right|
=O\left(c_nf^{-1}\left(\frac{k_{2,n}}{\rho_n n}\right)^{\alpha_0}\right)\nonumber\\+O_{a.co.}\left(c_n\sqrt{\frac{\log n}{nf\left(\lambda f^{-1}(\rho_n k_{1,n}/n)\right)}}\right) \label{PKNN3}
\end{eqnarray}

\noindent In addition, we get from conditions (\ref{h_g2})  and  (\ref{h_Phi_n}):
\begin{equation}
\max_{j=1,\dots,p}\left|g_{j,\theta_0}(\chi)\right|\left\lvert\left\lvert\pmb{\beta}-\pmb{\beta}_0\right\lvert\right\lvert=O(c_n),\label{PK4}
\end{equation}

\noindent and the claimed result is obtained from (\ref{PKNN1})-(\ref{PK4}), and because $c_n\rightarrow 0$ as $n\rightarrow\infty$.

\subsection{Proof of Theorem \ref{theorem2}}
The proof is the same, using Theorem 3.3(a)  rather than Theorem 3.3(b) in \cite{novo_2019}.

\end{document}